\documentclass[conference]{IEEEtran}
\usepackage{ifpdf}
\usepackage{cite}
\usepackage[cmex10]{amsmath}
\usepackage{array}
\usepackage{amsfonts}
\usepackage{mathrsfs}
\usepackage{arydshln}
\usepackage{slashbox}
\usepackage{graphicx}
\usepackage{float}
\usepackage{subfigure}
\usepackage{amssymb}
\usepackage{amsmath}
\usepackage{multirow}
\usepackage{multicol}
\usepackage{cite}
\usepackage[subfigure]{graphfig}
\usepackage{xcolor}
\usepackage{setspace}
\newcommand{\mv}[1]{\mbox{\boldmath{$ #1 $}}}
\usepackage[linesnumbered,ruled]{algorithm2e}

\newtheorem{lemma}{\underline{Lemma}}
\newtheorem{proposition}{\underline{Proposition}}

\newcommand{\qed}{\nobreak \ifvmode \relax \else
      \ifdim\lastskip<1.5em \hskip-\lastskip
      \hskip1.5em plus0em minus0.5em \fi \nobreak
      \vrule height0.75em width0.5em depth0.25em\fi}

\begin{document}
\title{Cellular-Enabled UAV Communication: Trajectory Optimization Under Connectivity Constraint}
\author{\IEEEauthorblockN{Shuowen~Zhang, Yong~Zeng, and Rui~Zhang}
\IEEEauthorblockA{ECE Department, National University of Singapore. Email: \{elezhsh,elezeng,elezhang\}@nus.edu.sg}}
\maketitle
\begin{abstract}
In this paper, we study a cellular-enabled unmanned aerial vehicle (UAV) communication system consisting of one UAV and multiple ground base stations (GBSs). The UAV has a mission of flying from an initial location to a final location, during which it needs to maintain reliable wireless connection with the cellular network by associating with one of the GBSs at each time instant. We aim to minimize the UAV mission completion time by optimizing its trajectory, subject to a quality of connectivity constraint of the GBS-UAV link specified by a minimum received signal-to-noise ratio (SNR) target, which needs to be satisfied throughout the mission. This problem is non-convex and difficult to be optimally solved. We first propose an effective approach to check its feasibility based on graph connectivity verification. Then, by examining the GBS-UAV association sequence during the UAV mission, we obtain useful insights on the optimal UAV trajectory, based on which an efficient algorithm is proposed to find an approximate solution to the trajectory optimization problem by leveraging techniques in convex optimization and graph theory. Numerical results show that our proposed {\hbox{trajectory design achieves near-optimal performance.}}
\end{abstract}
\vspace{-1mm}
\section{Introduction}
\vspace{-1mm}
The demand for unmanned aerial vehicles (UAVs), or drones, is expected to skyrocket in the near future, due to the continuous cost reduction in UAV manufacturing and the emergence of various new UAV-enabled applications in e.g., traffic control, cargo delivery, surveillance, aerial inspection, rescue and search, and communication platform \cite{survey}. It has been projected that the number of UAVs worldwide will approach at least 250,000 by the year 2035, in which over 175,000 will be used in commercial applications \cite{demand}. To practically realize the large-scale deployment of UAVs, it is of paramount importance to ensure that all UAVs can operate safely, which requires ultra-reliable, low-latency, and secure communication links between the UAV and the ground control stations (GCSs) for supporting the critical control and non-payload communications (CNPC) \cite{CNPC}. However, at present, almost all UAVs in the market rely on the simple direct point-to-point communication with their ground pilots over the unlicensed spectrum (e.g., ISM 2.4GHz), which is typically of limited data rate, unreliable, insecure, vulnerable to interference, and can only {\hbox{operate within the visual line of sight (LoS) range.}}

\begin{figure}[t]
  \centering
  \includegraphics[width=7cm]{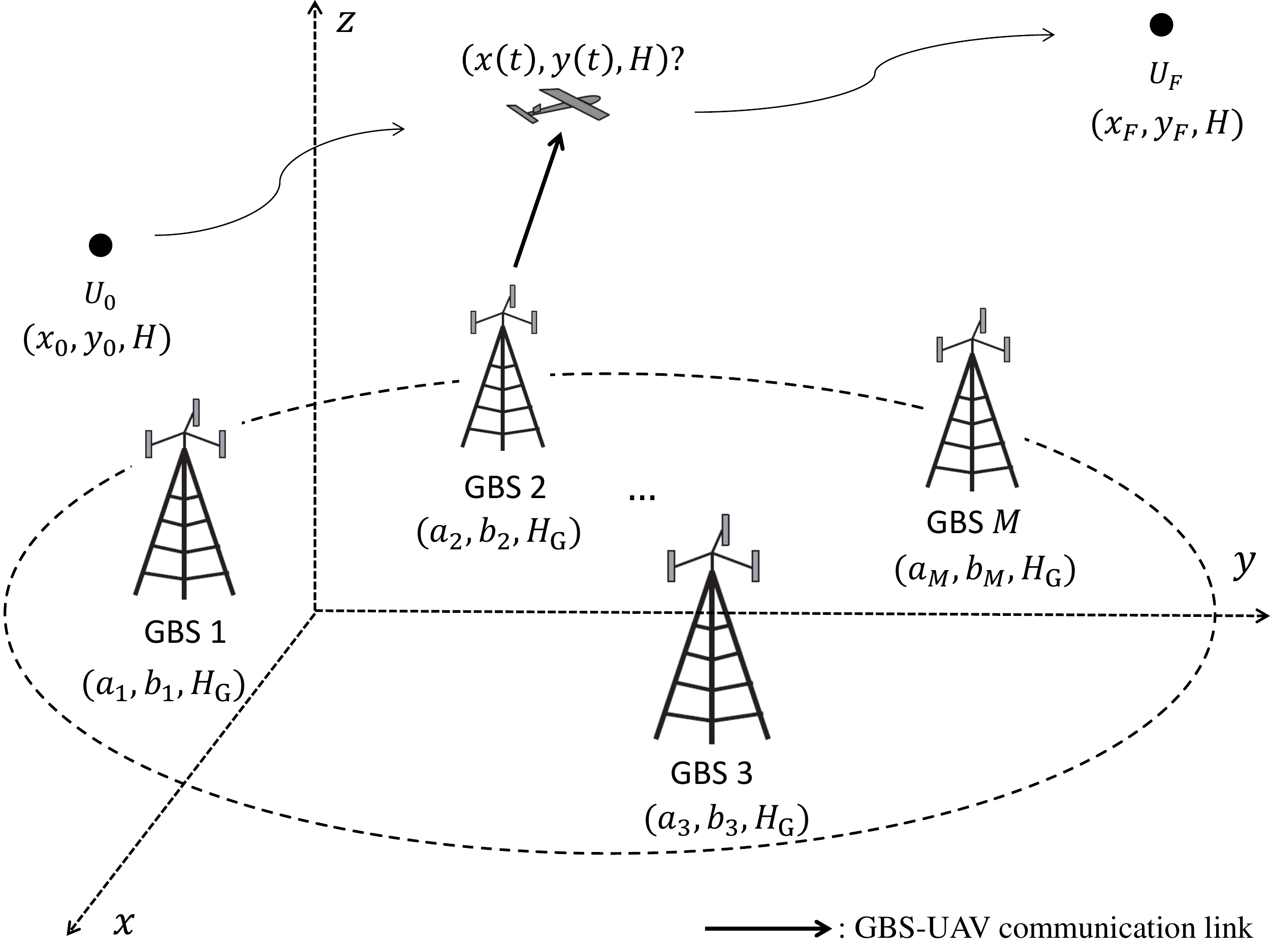}
  \vspace{-3mm}
  \caption{Illustration of a cellular-enabled UAV communication system.}\label{cellularUAV}
  \vspace{-8mm}
\end{figure}

In this paper, we consider a new and promising approach for realizing high-performance UAV-ground communication, namely \emph{cellular-enabled UAV communication}, as illustrated in Fig. \ref{cellularUAV}, where ground base stations (GBSs) in the existing 4G (fourth-generation) LTE (Long Term Evolution) or the forthcoming 5G (fifth-generation) cellular networks are utilized to enable communications between the UAVs and their ground users. Thanks to the almost ubiquitous accessibility worldwide and superior performance of today's LTE and future 5G wireless networks, cellular-enabled UAV communications are expected to achieve orders-of-magnitude performance improvement over the existing point-to-point UAV-ground communications, in terms of various performance metrics such as reliability, security, coverage and throughput. In particular, it potentially enables the safe and reliable CNPC links with flying UAVs for beyond LoS (BLoS) operations, which significantly extends the UAV operational range. Preliminary measurement results in industry and academia have demonstrated the feasibility of {\hbox{supporting UAVs using LTE networks \cite{LTEsky,Qualcomm}.}}

Despite its promising performance, many new design challenges need to be tackled for cellular-enabled UAV communications. Particularly, the UAV trajectory needs to be carefully designed such that the UAV can fulfill its mission (e.g., travelling between a pair of locations before a specified deadline) while at the same time meeting the communication requirements (for e.g., critical control by ground pilot) along its entire trajectory. Note that the trajectory design for cellular-enabled UAV communications is significantly different from that for UAV-enabled/aided wireless communication systems \cite{survey,relay,cyclical,energy}, where the UAV is employed as an \emph{aerial communication platform} (e.g., mobile relay \cite{relay} or aerial base station \cite{cyclical,energy}) to provide/enhance communication service to ground users. In this scenario, the UAV trajectory is designed for optimizing the performance of the ground users \cite{survey,relay,cyclical,energy}, in contrast to the case of cellular-enabled UAV communication considered in this paper where the UAV needs to be optimally served by the GBSs as an \emph{aerial user}. Moreover, as an aerial platform for serving ground communications, the UAV's trajectory can be in general more flexibly designed for the purpose of communication performance optimization only, as compared to the case considered in this paper as an aerial user which usually has its own mission for other applications (e.g., cargo delivery, video surveillance), thus imposing additional constraints on the trajectory planning. To the best of our knowledge, trajectory design for cellular-enabled UAV communications has not been studied in the literature, which motivates this work.

We investigate in this paper a basic cellular-enabled UAV communication system with one UAV and multiple GBSs, as shown in Fig. \ref{cellularUAV}. The UAV has a mission of flying between a given pair of initial and final locations, while maintaining its wireless connectivity with one of the GBSs at each time instant. We consider delay-limited communication between the UAV and the cellular network, where the quality of connectivity constraint is specified by a minimum received signal-to-noise ratio (SNR) requirement. In practice, one typical scenario is the command and control signal transmission from the remote pilot via the cellular network to the UAV. Under this setup, our objective is to minimize the UAV mission completion time by optimizing the UAV trajectory, subject to the minimum SNR requirement and a maximum speed constraint of the UAV. The formulated problem is non-convex and difficult to solve in general. First, we propose an efficient algorithm to check its feasibility based on graph connectivity. Then, by examining the GBS-UAV association sequence throughout the UAV mission, we reveal useful insights on the structure of the optimal UAV trajectory, based on which the problem is equivalently reformulated into a more tractable form. Finally, an efficient algorithm is proposed to find an approximate solution by leveraging convex optimization techniques and shortest-path algorithm in graph theory. Numerical results show that our proposed trajectory design achieves close performance to the optimal trajectory obtained via exhaustive search, yet with significantly reduced complexity.
\vspace{-1.5mm}
\section{System Model and Problem Formulation}
\vspace{-1.5mm}
As shown in Fig. \ref{cellularUAV}, we consider a cellular-enabled UAV communication system with $M>1$ GBSs and a UAV flying at a constant altitude of $H$ meters (m). We assume that the UAV has a mission of flying from an initial location $U_0$ to a final location $U_F$, while maintaining its wireless connection with the cellular network for communication. With a three-dimensional Cartesian coordinate system, we denote $(a_m,b_m,H_{\mathrm{G}})$ as the coordinate of the $m$th GBS, where for simplicity we assume that all the $M$ GBSs have the same altitude $H_{\mathrm{G}}$; denote $(x_0,y_0,H)$ and $(x_F,y_F,H)$ as the coordinates of $U_0$ and $U_F$, respectively; and denote $(x(t),y(t),H)$, $0\leq t\leq T$ as the time-varying coordinate of the UAV, where $T$ denotes the mission completion time. We further define ${\mv{g}}_m=[a_m,b_m]^T$, ${\mv{u}}_0=[x_0,y_0]^T$, ${\mv{u}}_F=[x_F,y_F]^T$ and ${\mv{u}}(t)=[x(t),y(t)]^T$ to represent the above locations projected on the horizontal ground plane, respectively, where ${\mv{u}}(0)={\mv{u}}_0$ and ${\mv{u}}(T)={\mv{u}}_F$.

For the purpose of exposition, we assume that the UAV as well as each GBS is equipped with a single antenna with omnidirectional unit gain, and the channel between the UAV and GBS is dominated by the LoS link, where the Doppler effect due to the UAV mobility is assumed to be compensated perfectly. The time-varying distance between the $m$th GBS and the UAV can be expressed as
\vspace{-1.5mm}\begin{equation}\label{distance}
d_{m}(t)=\sqrt{(H-H_{\mathrm{G}})^2+\|{\mv{u}}(t)-{\mv{g}}_{m}\|^2},\quad m\in \mathcal{M},
\vspace{-1.5mm}\end{equation}
where $\!\|\cdot\|\!$ denotes the Euclidean norm, and $\!\mathcal{M}\!=\!\{1,...,M\}\!$ denotes the set of GBSs. Let $h_m(t)\!\in\! \mathbb{C}$ denote the time-varying channel coefficient from the $m$th GBS to the UAV. It follows from (\ref{distance}) that the channel power gain {\hbox{can be modeled as}}
\vspace{-1.5mm}\begin{equation}\label{channel}
\!\!|h_{m}(t)|^2\!=\!\frac{{\beta_0}}{d_m^2(t)}\!=\!\frac{{\beta_0}}{{\!(H\!-\!H_{\mathrm{G}})^2+\|{\mv{u}}(t)\!-\!{\mv{g}}_m\|^2\!}},m\in \mathcal{M},
\vspace{-1.5mm}\end{equation}
where $\beta_0$ denotes the channel power gain at the reference distance of $d_0=1$m.
\vspace{-0.5mm}

We consider that at each time instant $t$ during the UAV mission, one single GBS indexed by $I(t)\!\in\!\mathcal{M}$ is selected to communicate with the UAV. In this paper, we focus on the scenario of downlink transmission from the GBS to the UAV, as illustrated in Fig. \ref{cellularUAV}; while our results are also applicable to uplink transmission from the UAV to the GBS. It can be observed from (\ref{channel}) that to maximize the received signal power at the UAV, the GBS closest to the UAV, namely, the one with the largest channel power gain, should be selected for communication with the UAV, i.e., $I(t)\!=\!\arg\!\underset{m\in \mathcal{M}}{\min}\|{\mv{u}}(t)-{\mv{g}}_m\|,0\!\leq\! t\!\leq\! T$. Consequently, the SNR at {\hbox{the UAV receiver is given by}}
\vspace{-2mm}\begin{equation}
\rho(t)=\frac{\gamma_0}{(H-H_{\mathrm{G}})^2+\underset{m\in \mathcal{M}}{\min}\ \|{\mv{u}}(t)-{\mv{g}}_m\|^2},\ 0\leq t\leq T,\label{rho}
\vspace{-3mm}\end{equation}
where $\gamma_0=\frac{P\beta_0}{\sigma^2}$ denotes the reference SNR, with $P$ and $\sigma^2$ denoting the transmission power of each GBS and the noise power at the UAV receiver, respectively. In this paper, the receiver SNR $\rho(t)$ is taken as the \emph{quality of connectivity} of the cellular-UAV communication link. We consider delay-limited communication for the UAV, where a minimum SNR target $\bar{\rho}$ needs to be satisfied at any time instant of the UAV mission. Notice from (\ref{rho}) that $\rho(t)$ is determined by the UAV trajectory ${\mv{u}}(t)$, which needs to be designed to satisfy the given SNR constraint on $\rho(t)$ for all $t\in [0,T]$.
\vspace{-0.5mm}

We aim to minimize the UAV mission completion time $T$ by optimizing the UAV trajectory ${\mv{u}}(t)$, subject to the minimum SNR constraint given by $\rho(t)\geq \bar{\rho},\ 0\leq t\leq T$. Furthermore, denote by $V_{\max}>0$ the maximum UAV speed. We thus have the constraint $\|\dot{\mv{u}}(t)\|\leq V_{\max},\ 0\leq t\leq T$, where $\dot{\mv{u}}(t)$ denotes the time-derivative of ${\mv{u}}(t)$. By explicitly expressing $\rho(t)$ according to (\ref{rho}), the minimum SNR constraint can be shown to be satisfied if and only if the horizontal distance between the UAV and its closest GBS, $\underset{m\in\mathcal{M}}{\min}\ \|{\mv{u}}(t)-{\mv{g}}_m\|$, is no larger than $\bar{d}\overset{\Delta}{=}\sqrt{\tfrac{\gamma_0}{\bar{\rho}}-(H-H_\mathrm{G})^2}$ at any time instant during the UAV mission. Note that a smaller $\bar{d}$ corresponds to a larger SNR target $\bar{\rho}$, and hence more stringent constraint on the quality of connectivity. The optimization problem is formulated as follows:
\vspace{-2.5mm}
\begin{align}
\mbox{(P1)}\quad \underset{T,{\mv{u}}(t)}{\min}\quad &T\\[-1mm]
\mathrm{s.t.}\quad & {\mv{u}}(0)={\mv{u}}_0\label{P1c_i}\\[-0.5mm]
& {\mv{u}}(T)={\mv{u}}_F\label{P1c_f}\\[-0.5mm]
& \underset{m\in \mathcal{M}}{\min}\ \|{\mv{u}}(t)-{\mv{g}}_m\|\leq \bar{d},\quad 0\leq t\leq T\label{P1c_SNR}\\[-0.5mm]
& \|\dot{\mv{u}}(t)\|\leq V_{\max},\quad 0\leq t\leq T.\label{P1c_v}
\end{align}

Note that Problem (P1) is a non-convex optimization problem, since the left-hand side (LHS) of the constraint in (\ref{P1c_SNR}) is the pointwise minimum of a set of convex functions, thus being a non-convex function in general. Moreover, ${\mv{u}}(t)$ is a continuous function of $t$, thus Problem (P1) essentially involves an infinite number of optimization variables. Therefore, the optimal solution to Problem (P1) is in general challenging to obtain. In the following sections, we first check the feasibility of Problem (P1), and then propose an efficient approach for finding an approximate solution if it is feasible, based on graph theory and convex optimization techniques.
\vspace{-1mm}
\section{Feasibility of Problem (P1): Graph Connectivity Based Verification}\label{sec_feasibility}
\vspace{-1mm}
In this section, we study the feasibility of Problem (P1). Notice that Problem (P1) is feasible if and only if it is feasible without the UAV speed constraint in (\ref{P1c_v}), since for any trajectory ${\mv{u}}(t)$ that satisfies the constraints in (\ref{P1c_i}), (\ref{P1c_f}) and (\ref{P1c_SNR}), we can always construct a feasible solution to Problem (P1) by letting the UAV travel in the same path as ${\mv{u}}(t)$ with maximum speed $V_{\max}$. Therefore, the feasibility of Problem (P1) can be checked by solving the following problem:
\begin{align}
\mbox{(P1-F)}\quad {\mathrm{Find}}\quad &T,{\mv{u}}(t)\\[-0.5mm]
\mathrm{s.t.}\quad & (\ref{P1c_i}), (\ref{P1c_f}), (\ref{P1c_SNR}).
\end{align}
Due to the non-convex constraint in (\ref{P1c_SNR}) and the continuous variable ${\mv{u}}(t)$, it is difficult to directly solve Problem (P1-F). In the following, we propose an efficient approach for solving Problem (P1-F) by examining the sequential GBS-UAV association during the UAV mission {\hbox{which is implied in (\ref{P1c_SNR}).}}

Specifically, notice that with any given UAV trajectory ${\mv{u}}(t)$, the constraint in (\ref{P1c_SNR}) is satisfied if and only if there exists a sequence of GBSs that are successively associated with the UAV over the time horizon $[0,T]$, with the horizontal distance between the UAV and its associated GBS no greater than $\bar{d}$ at any time instant $t\in [0,T]$. We introduce an auxiliary vector ${\mv{I}}=[I_1,...,I_N]^T$ with $I_i\in \mathcal{M},\ \forall i$ to represent the \emph{GBS-UAV association sequence}, which indicates that the UAV is first associated with GBS $I_1$, and then handed over to GBS $I_2$ after a certain amount of time, etc., with $N-1$ denoting the total number of \emph{GBS handovers}. We then have the following proposition.
\begin{proposition}\label{prop_feas}
Problem (P1) is feasible if and only if there exists a GBS-UAV association sequence ${\mv{I}}=[I_1,...,I_N]^T$ that satisfies the following conditions:
\begin{align}
\|{\mv{u}}_0-{\mv{g}}_{I_1}\|&\leq\bar{d}\label{feas_i}\\[-0.5mm]
\|{\mv{u}}_F-{\mv{g}}_{I_N}\|&\leq \bar{d}\label{feas_f}\\[-0.5mm]
\|{\mv{g}}_{I_{i+1}}-{\mv{g}}_{I_i}\|&\leq2\bar{d},\quad i=1,...,N-1\label{feas_GBS}\\[-0.5mm]
I_i&\in\mathcal{M},\quad i=1,...,N.\label{feas_I}
\end{align}
\end{proposition}
\vspace{-1mm}
\begin{IEEEproof}
Please refer to Appendix A.
\end{IEEEproof}

\addtolength{\subfigcapskip}{-2mm}
\begin{figure}[t]
\centering
  \subfigure[Horizontal locations of $U_0$, $U_F$ and $M=10$ GBSs]{
  \includegraphics[width=8cm]{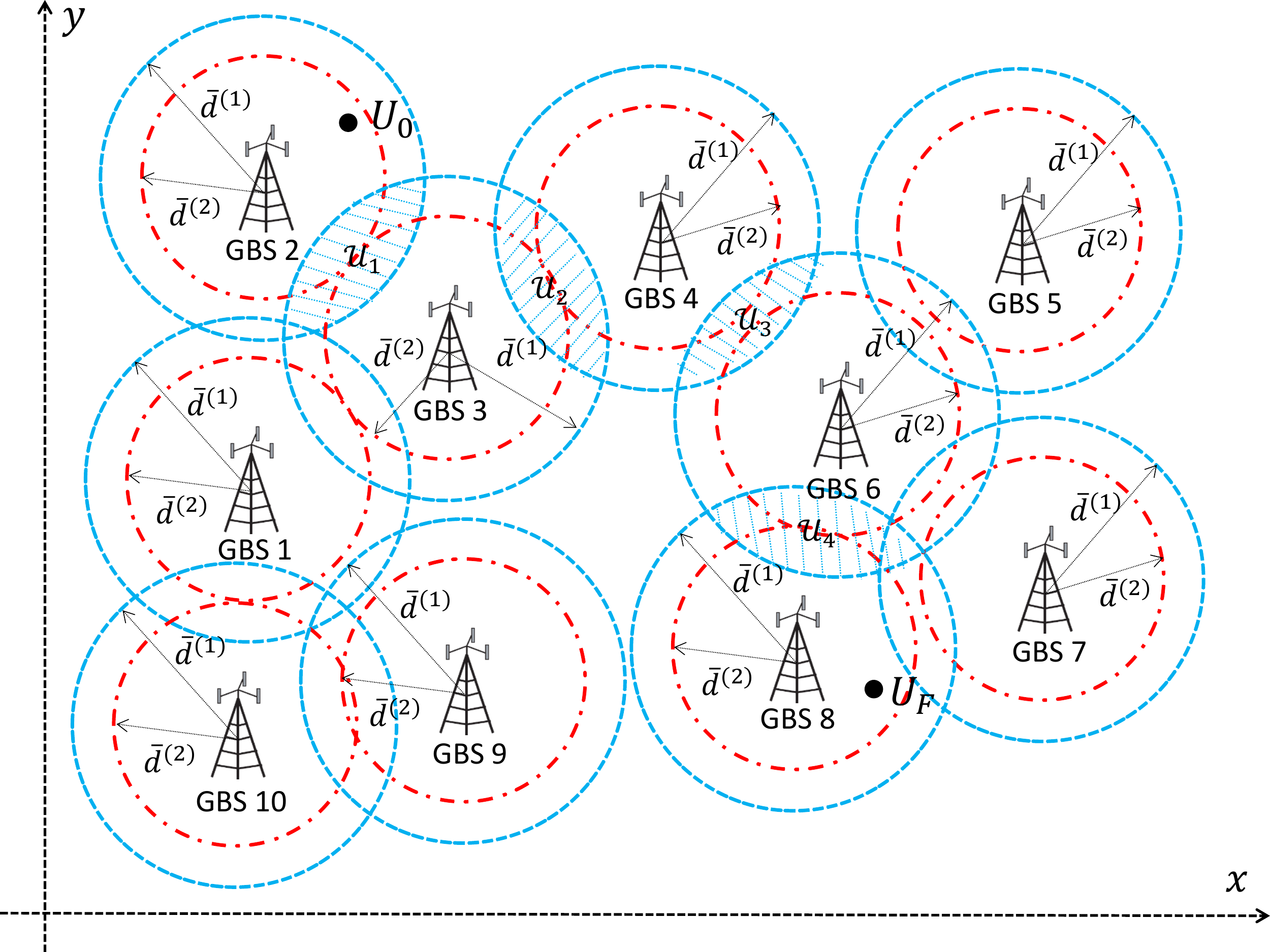}}
  \vspace{-2mm}

  \subfigure[Graph ${G}$ with $\bar{d}=\bar{d}^{(1)}$: Feasible case]{
  \includegraphics[width=3.8cm]{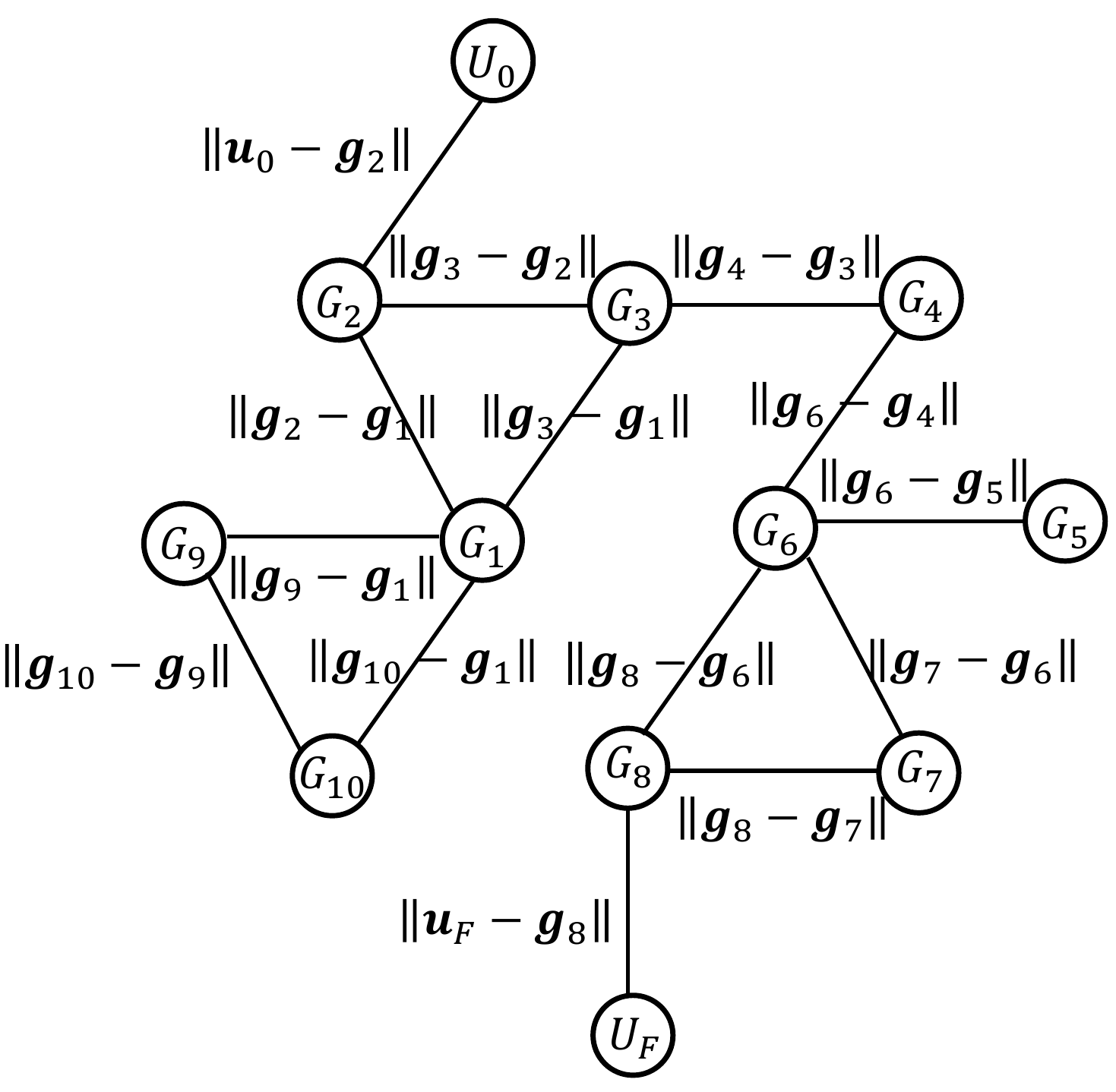}}
  \subfigure[Graph ${G}$ with $\bar{d}=\bar{d}^{(2)}$: Infeasible case]{
  \includegraphics[width=3.8cm]{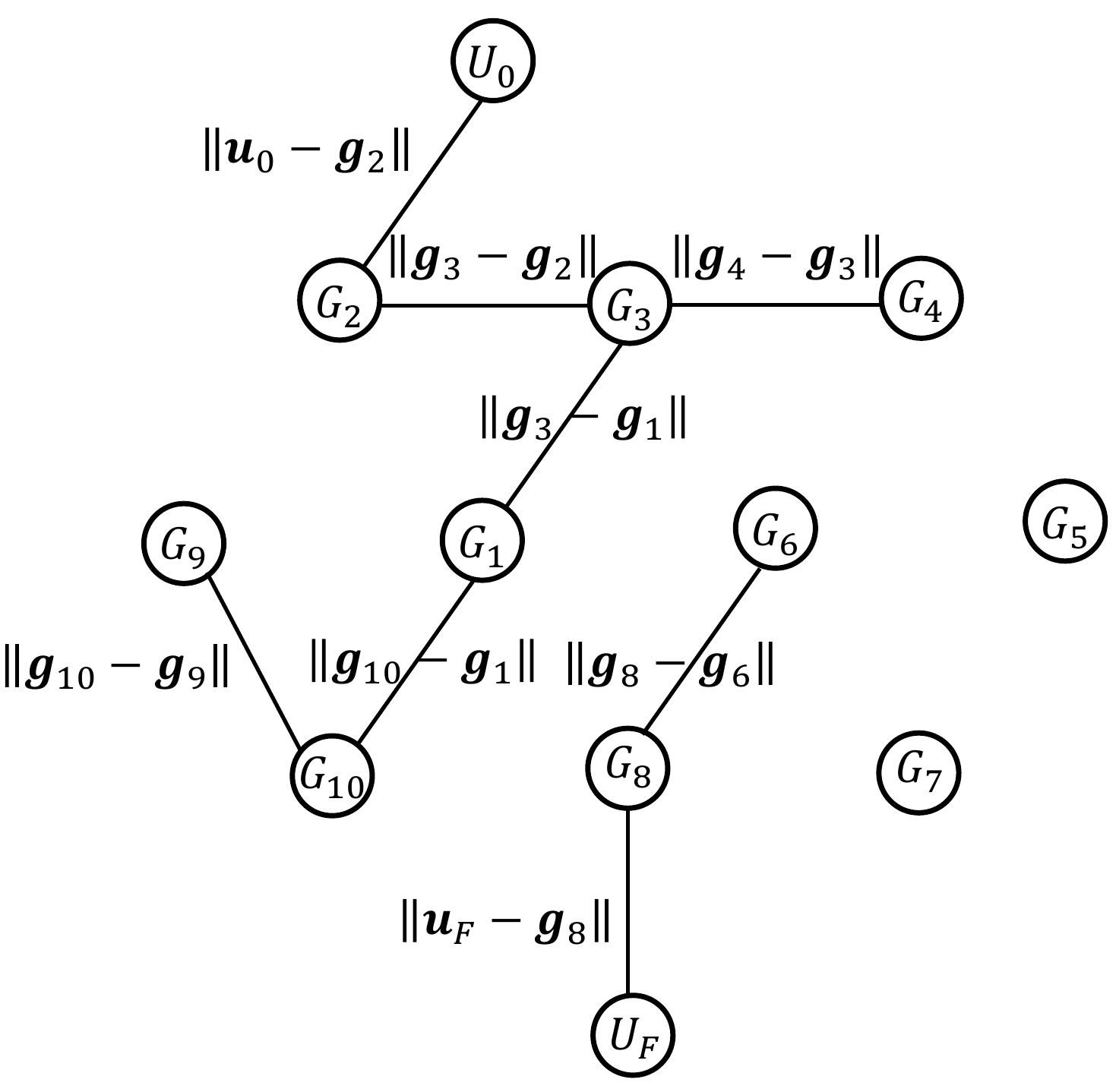}}
  \vspace{-2mm}
  \caption{Illustration of feasibility verification for Problem (P1) based on graph connectivity.}
  \label{fig_feas}
  \vspace{-8mm}
\end{figure}

Based on the results in Proposition \ref{prop_feas}, the feasibility of Problem (P1) can be checked via the following procedure by leveraging \emph{graph connectivity}. First, we construct an undirected weighted \emph{graph} denoted by ${G}=(V,E)$, where the vertex set $V$ is given by
\vspace{-0.5mm}\begin{equation}\label{vertex}
V=\{U_0,G_1,...,G_M,U_F\},
\vspace{-1mm}\end{equation}
where $U_0$ and $U_F$ represent the UAV initial and final locations, respectively, and $G_m$ represents the $m$th GBS; the edge set $E$ is given by
\begin{align}\label{edge}
E=&\{(U_0,G_m):\|{\mv{u}}_0-\!{\mv{g}}_m\|\leq \bar{d},\ m\in \mathcal{M}\}\nonumber\\[-0.5mm]
\cup&\{(G_m,G_n):\|{\mv{g}}_m-{\mv{g}}_n\|\leq 2\bar{d},\ m,n\in \mathcal{M},m\neq n\}\nonumber\\[-0.5mm]
\cup&\{(U_F,G_m):\|{\mv{u}}_F-{\mv{g}}_m\|\leq \bar{d},\ m\in \mathcal{M}\}.
\end{align}
The weight of each edge is given by
\begin{align}\label{weight}
&W(U_0,G_m)=\|{\mv{u}}_0-{\mv{g}}_m\|,\ W(U_F,G_m)=\|{\mv{u}}_F-{\mv{g}}_m\|\nonumber\\[-0.5mm]
&W(G_m,G_n)=\|{\mv{g}}_m-{\mv{g}}_n\|,\quad m,n\in \mathcal{M},m\neq n.
\end{align}
Note that an edge $(U_0,G_m)$ or $(U_F,G_m)$ exists if and only if the horizontal distance between $U_0$ or $U_F$ with the $m$th GBS is no larger than $\bar{d}$, respectively, whose weight represents this distance; an edge $(G_m,G_n)$ exists if and only if the distance between the $m$th and the $n$th GBSs is no larger than $2\bar{d}$, whose weight represents their distance. For illustration, we consider an example of a system with horizontal locations of $U_0$, $U_F$ and $M=10$ GBSs shown in Fig. \ref{fig_feas}(a). We show in Fig. \ref{fig_feas}(b) and Fig. \ref{fig_feas}(c) the construction of the graph $G$ with two different values of $\bar{d}$ given by $\bar{d}^{(1)}$ and $\bar{d}^{(2)}=\frac{3}{4}\bar{d}^{(1)}$, respectively, which are also illustrated in Fig. \ref{fig_feas}(a).

It then follows from Proposition \ref{prop_feas} and the definition of ${G}$ that Problem (P1) is feasible if and only if $U_0$ and $U_F$ are \emph{connected}, i.e., ${G}$ contains a \emph{path} from $U_0$ to $U_F$ \cite{graph}. The connectivity between $U_0$ and $U_F$ can be readily verified via various existing algorithms, e.g., breadth-first search, with time complexity $\mathcal{O}(M+2)$ \cite{graph}, where $\mathcal{O}(\cdot)$ denotes the standard big-O notation. Note that construction of the graph ${G}$ requires time complexity $\mathcal{O}\big(2M+\frac{M(M-1)}{2}\big)$. Thus, the total time complexity for checking the feasibility of Problem (P1) is $\mathcal{O}\big(\frac{M^2}{2}+\frac{5M}{2}+2\big)$. As an example, it can be observed that $U_0$ and $U_F$ are connected in the graph shown in Fig. \ref{fig_feas}(b) (e.g., a path can be easily found as $(U_0,G_2,G_3,G_4,G_6,G_8,U_F)$, which corresponds to a GBS-UAV association sequence ${\mv{I}}=[2,3,4,6,8]^T$), thus Problem (P1) is feasible with $\bar{d}=\bar{d}^{(1)}$; on the other hand, $U_0$ and $U_F$ are not connected in the graph shown in Fig. \ref{fig_feas}(c), namely, Problem (P1) is infeasible with $\bar{d}=\bar{d}^{(2)}=\frac{3}{4}\bar{d}^{(1)}$, due to the more stringent constraint in (\ref{P1c_SNR}) with a smaller $\bar{d}$.

\section{Proposed Solution to Problem (P1)}\label{sec_solution}
In this section, we propose an efficient algorithm for finding an approximate solution to Problem (P1) if it is verified to be feasible. Specifically, by leveraging the GBS-UAV association established in Section \ref{sec_feasibility} and by exploiting the special structure of the optimal UAV trajectory, we equivalently transform Problem (P1) into a joint optimization problem of the GBS-UAV association and handover locations of the UAV. A low-complexity algorithm is then proposed for this problem by leveraging convex optimization and \hbox{graph theory.}
\subsection{Problem Reformulation Based on GBS-UAV Association}
To start with, we reformulate Problem (P1) into a more tractable form by re-expressing the constraint in (\ref{P1c_SNR}) via explicitly characterizing the GBS-UAV association during the UAV mission. Specifically, recall from Section \ref{sec_feasibility} that the indices of the GBSs sequentially associated with the UAV can be represented by an auxiliary vector ${\mv{I}}=[I_1,...,I_N]^T$, where $I_i\in \mathcal{M}$ denotes the GBS associated with the UAV between the $(i-1)$th and the $i$th handovers. We further introduce a set of auxiliary variables $\{T_i\}_{i=1}^{N}$, where $T_i$ denotes the time duration between the $(i-1)$th and $i$th handovers for $i=2,...,N-1$, $T_1$ denotes the time duration from the mission start to the first handover, and $T_{N}$ denotes the time duration from the $(N-1)$th handover to the mission completion. By leveraging the auxiliary variables ${\mv{I}}$ and $\{T_i\}_{i=1}^{N}$, we provide the following proposition.
\begin{proposition}\label{prop_P2eq}
Problem (P1) is equivalent to the following problem:
\vspace{-2mm}
\begin{align}
{\mbox{(P2)}}\underset{T,{\mv{u}}(t),{\mv{I}},\{T_i\}_{i=1}^N}{\min} & T\\[-1mm]
\mathrm{s.t.}\quad & (\ref{P1c_i}),(\ref{P1c_f}),(\ref{P1c_v}),(\ref{feas_i}),(\ref{feas_f}),(\ref{feas_GBS}),(\ref{feas_I})\\[-1mm]
& \|{\mv{u}}(t)-{\mv{g}}_{I_i}\|\leq \bar{d},\quad  \forall t\in\bigg[\sum_{j=1}^{i-1}T_j,\sum_{j=1}^i T_j\bigg], \nonumber\\[-1mm]
& \qquad\qquad\qquad\qquad\quad i=1,...,N\label{P2c_SNR}\\[-1mm]
& \sum_{i=1}^N T_i=T.\label{P2c_t}
\end{align}
\end{proposition}
\vspace{-1mm}
\begin{IEEEproof}
Please refer to Appendix B.
\end{IEEEproof}

For convenience of exposition, we define the horizontal location of the UAV where it is handed over from GBS $I_{i}$ to GBS $I_{i+1}$, i.e., at the $i$th handover point, as
\vspace{-2mm}\begin{equation}
{{\mv{u}}^i}\overset{\Delta}{=}{\mv{u}}\bigg(\sum_{j=1}^i T_j\bigg),\quad i=1,...,N-1.
\vspace{-2mm}\end{equation}
Note that under the constraints in (\ref{P2c_SNR}), each $i$th handover point has a horizontal distance no larger than $\bar{d}$ with both GBSs $I_i$ and $I_{i+1}$, i.e., the feasible region of ${\mv{u}}^i$ is given by
\begin{align}\label{Ui}
&\mathcal{U}_i=\{{\mv{u}}^i\in \mathbb{R}^{2\times 1}:\|{\mv{u}}^i-{\mv{g}}_{I_i}\|\leq \bar{d},\ \|{\mv{u}}^{i}-{\mv{g}}_{I_{i+1}}\|\leq \bar{d}\},\nonumber\\[-0.5mm]
& \qquad\qquad\qquad\qquad\qquad\qquad\qquad\quad i=1,...,N-1.
\end{align}
In Fig. \ref{fig_feas}(a), we illustrate $\mathcal{U}_i$'s by taking the example of $\bar{d}=\bar{d}^{(1)}$ and ${\mv{I}}=[2,3,4,6,8]^T$. For consistence, we further define ${{\mv{u}}^0}\overset{\Delta}{=}{{\mv{u}}(0)}={\mv{u}}_0$ and ${\mv{u}}^N\overset{\Delta}{=}{\mv{u}}(T)={\mv{u}}_F$ as the horizontal locations of the $0$th and the $N$th handover points, respectively.
\subsection{Structure of the Optimal UAV Trajectory}
Next, based on the UAV handover locations $\{{\mv{u}}^i\}_{i=0}^N$ defined above, we are ready to present the following proposition.
\begin{proposition}\label{prop_P2}
The optimal solution to Problem (P2) satisfies the following conditions:
\vspace{-1mm}
\begin{align}
T_i&=\frac{\left\|{{\mv{u}}^i}-{{\mv{u}}^{i-1}}\right\|}{V_{\max}},\quad i=1,...,N\label{Ti}\\[-1mm]
{\mv{u}}(t)&={{\mv{u}}^{i-1}}+\bigg(t-\sum_{j=1}^{i-1} T_j\bigg)V_{\max}\frac{{\mv{u}}^{i}-{\mv{u}}^{i-1}}{\|{\mv{u}}^{i}-{\mv{u}}^{i-1}\|},\nonumber\\[-1mm]
&\qquad\qquad\quad\quad t\in\bigg[\sum_{j=1}^{i-1}T_j,\sum_{j=1}^i T_j\bigg],\ i=1,...,N\label{ut}\\[-1mm]
T&=\sum_{i=1}^N\frac{\left\|{{\mv{u}}^i}-{{\mv{u}}^{i-1}}\right\|}{V_{\max}}.\label{T}
\end{align}
\end{proposition}
\vspace{-1mm}
\begin{IEEEproof}
Please refer to Appendix C.
\end{IEEEproof}
The results in Proposition \ref{prop_P2} indicate that with the optimal solution to Problem (P2) as well as Problem (P1), the UAV should fly from $U_0$ to $U_F$ by following a path consisting of \emph{connected line segments} with the \emph{maximum speed}. Moreover, the UAV is associated with the same GBS while it flies within each line segment, and the starting and ending points of each $i$th segment are the $(i-1)$th and the $i$th handover points with horizontal locations specified by ${\mv{u}}^{i-1}$ and ${\mv{u}}^{i}$, respectively.

By leveraging this optimal structure, Problem (P2) can be readily shown to be equivalent to the following problem, which aims to minimize the \emph{total flying distance} of the UAV by jointly optimizing the \emph{GBS-UAV association sequence} ${\mv{I}}$ and the \emph{handover locations} $\{{\mv{u}}^i\}_{i=0}^N$:
\vspace{-2mm}
\begin{align}
{\mbox{(P3)}}\quad \underset{{\mv{I}},\{{\mv{u}}^i\}_{i=0}^N}{\min}\quad & \sum_{i=1}^N \|{\mv{u}}^i-{\mv{u}}^{i-1}\|\\[-0.5mm]
\mathrm{s.t.}\quad & {\mv{u}}^0={\mv{u}}_0\label{P3c_i}\\[-0.5mm]
& {\mv{u}}^N={\mv{u}}_F\label{P3c_f}\\[-0.5mm]
& \|{\mv{u}}^i-{\mv{g}}_{I_i}\|\leq \bar{d},\quad i=1,...,N \label{P3c_SNR1}\\[-0.5mm]
& \|{\mv{u}}^{i-1}-{\mv{g}}_{I_{i}}\|\leq \bar{d},\quad i=1,...,N \label{P3c_SNR2}\\[-0.5mm]
& (\ref{feas_i}),(\ref{feas_f}),(\ref{feas_GBS}),(\ref{feas_I}).
\end{align}
Notice that by characterizing the continuous UAV trajectory ${\mv{u}}(t)$ with a discrete set of handover locations $\{{\mv{u}}^i\}_{i=0}^N$, Problem (P3) involves a significantly reduced number of optimization variables compared to Problem (P2). It is worth noting that due to the equivalence between Problem (P1) and Problem (P2) as shown in Proposition \ref{prop_P2eq}, Problem (P3) is equivalent to Problem (P1), whose optimal solution can be readily obtained by substituting the optimal solution to Problem (P3) into (\ref{ut}) and (\ref{T}). Thus, the remaining task is to solve Problem (P3).
\subsection{Proposed Solution to Problem (P3)}
Note that Problem (P3) is still a non-convex optimization problem due to the discrete variables $I_i$'s and the coupling of $I_i$'s and $\{{\mv{u}}^i\}_{i=0}^N$ through (\ref{P3c_SNR1}) and (\ref{P3c_SNR2}). Nevertheless, it is worth noting that with any given GBS-UAV association sequence $\mv{I}$, Problem (P3) is a convex optimization problem, since the feasible set of each $i$th UAV handover location, $\mathcal{U}_i$, is convex, as can be observed from Fig. \ref{fig_feas}(a). The optimal handover locations with given $\mv{I}$ denoted by $\{{{\mv{u}}^i}^\star({\mv{I}})\}_{i=0}^N$ can be thus efficiently obtained via existing software, e.g., CVX \cite{cvx}, with polynomial time complexity over $N$ \cite{convex}. Therefore, the optimal solution to Problem (P3) can be obtained by finding $\{{{\mv{u}}^i}^\star({\mv{I}})\}_{i=0}^N$ for all feasible solutions of $\mv{I}$, and selecting the one that yields the minimum objective value. To reduce the search space of $\mv{I}$, we provide the following lemma.
\begin{lemma}\label{lemma_I}
The optimal solution to Problem (P3) satisfies $I_i\neq I_j,\ \forall i\neq j$ and $N\leq M$.
\end{lemma}
\begin{IEEEproof}
Please refer to Appendix D.
\end{IEEEproof}
Lemma \ref{lemma_I} implies that the UAV shall not be associated with the same GBS in two non-consecutive time intervals, and the total number of handovers during the UAV mission is no larger than $M-1$. This is expected since in order to minimize the total flying distance, the UAV shall not return to the neighbourhood of its previously traveled paths.

\begin{figure}[t]
  \centering
  \includegraphics[width=7cm]{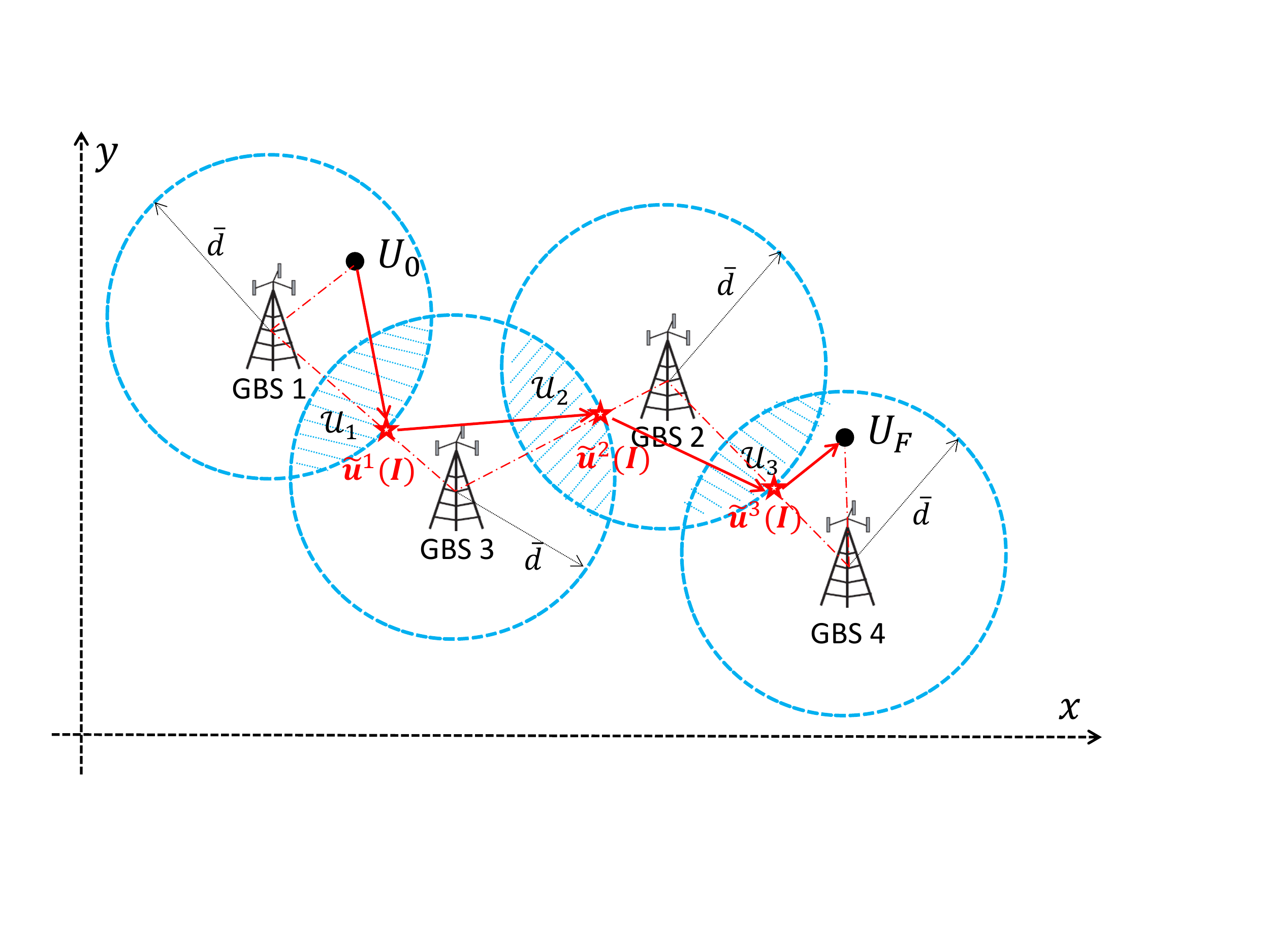}
  \vspace{-4mm}
  \caption{Illustration of a feasible solution of handover locations $\{\tilde{\mv{u}}^i({\mv{I}})\}_{i=1}^{N-1}$ with given ${\mv{I}}=[1,3,2,4]^T$.}\label{handover}
  \vspace{-7mm}
\end{figure}

Based on Lemma \ref{lemma_I}, the optimal GBS-UAV association for Problem (P3) can be found by solving the following problem:
\begin{align}
{\mbox{(P4)}}\quad \underset{{\mv{I}}}{\min}\quad & \sum_{i=1}^N \|{{\mv{u}}^i}^\star({\mv{I}})-{{\mv{u}}^{i-1}}^\star({\mv{I}})\|\\[-1mm]
\mathrm{s.t.}\quad & I_i\neq I_j, \quad  \forall i\neq j,\ i,j=1,...,N\label{P4c_different}\\[-1mm]
& (\ref{feas_i}),(\ref{feas_f}),(\ref{feas_GBS}),(\ref{feas_I}).
\end{align}
Note that optimally solving Problem (P4) via exhaustive search involves finding all possible paths from $U_0$ to $U_F$ in the graph $G=(V,E)$ defined in (\ref{vertex})-(\ref{weight}) in Section \ref{sec_feasibility}, which requires prohibitive time complexity (e.g., $\mathcal{O}((M+2)!)$ via depth-first search \cite{graph}) and may not be affordable even for moderate $M$. Hence, we aim to find an approximate solution to Problem (P4) instead by minimizing an upper bound of its objective function. Specifically, note that a feasible solution to Problem (P3) with given $\mv{I}$ can be obtained as $\{\tilde{\mv{u}}^i({\mv{I}})\}_{i=0}^N$, with $\tilde{\mv{u}}^0({\mv{I}})={\mv{u}}_0$, $\tilde{\mv{u}}^N({\mv{I}})={\mv{u}}_F$, and
\vspace{-1mm}
\begin{equation} \tilde{\mv{u}}^i({\mv{I}})={\mv{g}}_{I_i}+\bar{d}\frac{{\mv{g}}_{I_{i+1}}-{\mv{g}}_{I_{i}}}{\|{\mv{g}}_{I_{i+1}}-{\mv{g}}_{I_{i}}\|},\quad  i=1,...,N-1,
\vspace{-1mm}\end{equation}
where the horizontal location of each handover point lies on the line segment between the horizontal locations of its consecutively associated two GBSs and has distance $\bar{d}$ to the horizontal location of its formerly associated GBS, as illustrated in Fig. \ref{handover}. It then follows that
\vspace{-2mm}
\begin{align}\label{upp}
&\sum_{i=1}^N \|{{\mv{u}}^{i}}^{\star}({\mv{I}})-{{\mv{u}}^{i-1}}^{\star}({\mv{I}})\|\leq \sum_{i=1}^N \|{\tilde{\mv{u}}^{i}}({\mv{I}})-{\tilde{\mv{u}}^{i-1}}({\mv{I}})\|\nonumber\\[-2mm]
\overset{(a)}{\leq}& \|{\mv{u}}_0-{\mv{g}}_{I_1}\|+\sum_{i=1}^{N-1} \|{\mv{g}}_{I_{i+1}}-{\mv{g}}_{I_i}\|+\|{\mv{u}}_F-{\mv{g}}_{I_N}\|,
\end{align}
where $(a)$ can be derived by applying the triangle inequality, and is also illustrated in Fig. \ref{handover}. Note that finding the optimal ${\mv{I}}$ that minimizes the upper bound of the objective function of Problem (P4) given in (\ref{upp}) subject to the constraints in (\ref{P4c_different}) and (\ref{feas_i})-(\ref{feas_I}) can be shown to be equivalent to finding the \emph{shortest path} from $U_0$ to $U_F$ in the graph ${G}$, which can be efficiently obtained via various existing algorithms with low complexity, e.g., the Dijkstra algorithm with time complexity $\mathcal{O}((M+2)^2)$ \cite{graph}.
\vspace{-1mm}
\subsection{Overall Algorithm for Problem (P1)}
\vspace{-1mm}
To summarize, we provide Algorithm \ref{algo_P1} for Problem (P1), which first checks its feasibility; then provides an approximate solution $(\tilde{T},\tilde{\mv{u}}(t))$ if it is feasible, or sets $\tilde{T}=\infty$, $\tilde{\mv{u}}(t)={\mv{u}}_0$ otherwise. Note that for the case of Problem (P1) being feasible, the obtained $(\tilde{T},\tilde{\mv{u}}(t))$ is generally a suboptimal solution to Problem (P1), thus $\tilde{T}$ is in general an upper bound on its optimal value. It is also worth noting that Algorithm \ref{algo_P1} can be shown to be of polynomial time complexity over $M$.
\vspace{-4mm}
\begin{algorithm}[!htb]\label{algo_P1}
\caption{Proposed Algorithm for Problem (P1)}
\SetKwData{Index}{Index}
\KwIn{$\bar{d}$, ${\mv{u}}_0$, ${\mv{u}}_F$, $\{{\mv{g}}_m\}_{m=1}^{M}$, $H$, $H_G$, $V_{\max}$}
\KwOut{$\tilde{T}$, $\tilde{\mv{u}}(t)$}
Construct a graph ${G}=(V,E)$ based on (\ref{vertex})-(\ref{weight}).\\[-0.5mm]
\eIf{there exists a path from $U_0$ to $U_F$ in ${G}$}{\vspace{-0.5mm}
Find the shortest path from $U_0$ to $U_F$ in ${G}$ via Dijkstra algorithm, and denote the path as $(U_0,G_{\tilde{I}_1},...,G_{\tilde{I}_N},U_F)$. Obtain $\tilde{\mv{I}}=[\tilde{I}_1,...,\tilde{I}_N]^T$.\\
Obtain $\{{{\mv{u}}^i}^\star(\tilde{\mv{I}})\}_{i=0}^N$ by solving Problem (P3) with given $\tilde{\mv{I}}$ via convex optimization.\\
Obtain $\tilde{T}$ and $\tilde{\mv{u}}(t)$ by substituting $\{{{\mv{u}}^i}^\star(\tilde{\mv{I}})\}_{i=0}^N$ for $\{{\mv{u}}^i\}_{i=0}^N$ in (\ref{T}) and (\ref{ut}), respectively.}{\vspace{-0.5mm}Set $\tilde{T}=\infty$; $\tilde{\mv{u}}(t)={\mv{u}}_0,\  0\leq t\leq \tilde{T}$.}
\vspace{-0.5mm}\end{algorithm}
\section{Numerical Results}
In this section, we provide numerical results to evaluate the performance of our proposed trajectory design. We suppose that $M=11$ GBSs are uniformly distributed in a $10^4\mathrm{m}\times 10^4 \mathrm{m}$ (i.e., $10\mathrm{km}\times 10\mathrm{km}$) region. The UAV's initial and final locations projected on the horizontal plane are set as ${\mv{u}}_0=[2000,2000]^T$ and ${\mv{u}}_F=[8000,8000]^T$, respectively. The altitude of the UAV and each GBS is set as $H=90$m and $H_G=12.5$m, respectively. The maximum UAV speed is set as $V_{\max}=50$m/s. The reference SNR at distance $d_0=1$m is set as $\gamma_0=\frac{P\beta_0}{\sigma^2}=80$dB. For comparison, we consider the optimal UAV trajectory design for Problem (P1), which is obtained by solving Problem (P4) via exhaustive search over all feasible GBS-UAV associations. In addition, we consider the simple straight flight trajectory, where the UAV flies from $U_0$ to $U_F$ in a straight path with maximum speed.

\begin{figure}[!htb]
  \centering
  \includegraphics[width=8cm]{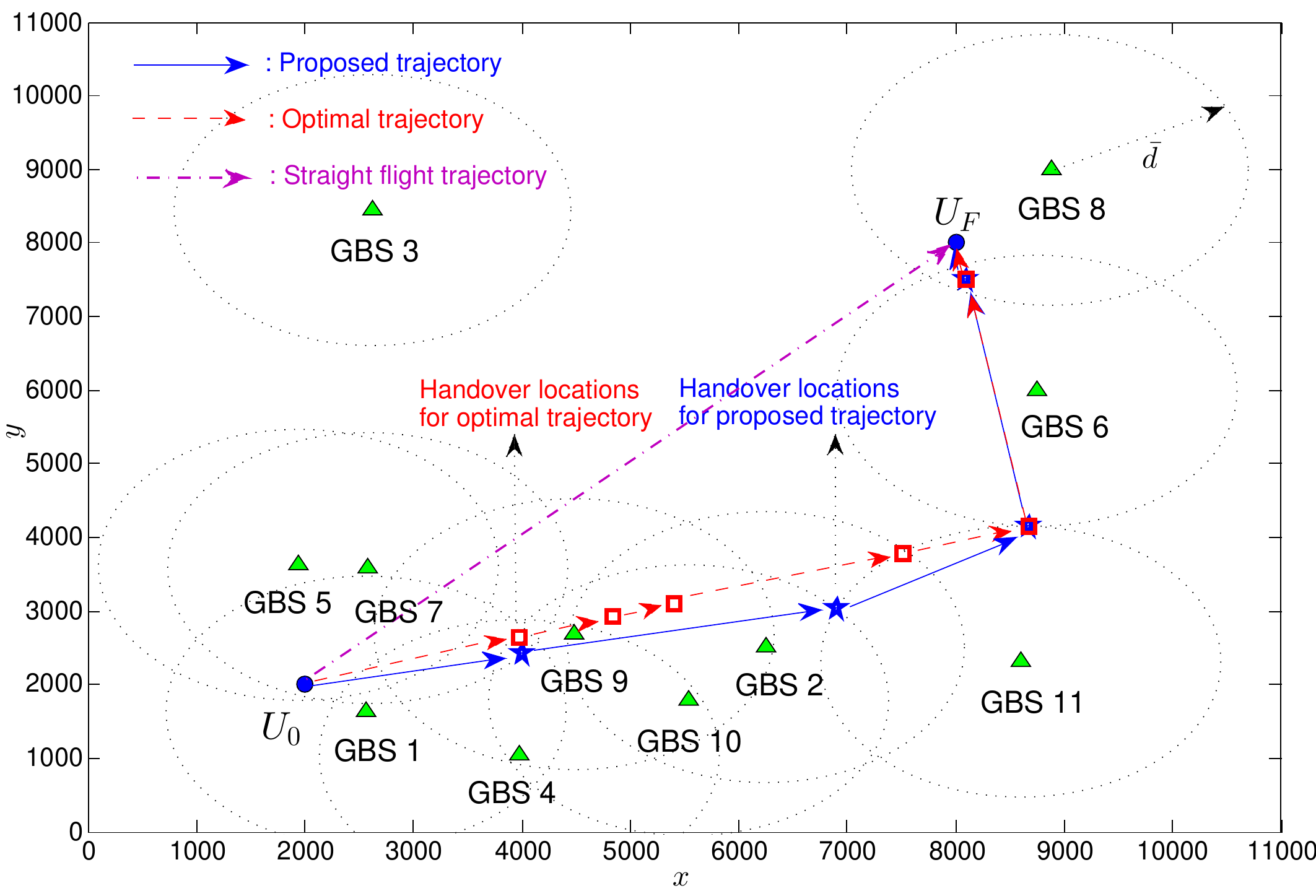}
  \vspace{-4mm}
  \caption{Illustration of UAV trajectory design with $\bar{\rho}=\bar{\rho}_{\max}$.}\label{trajectory}
  \vspace{-3mm}
\end{figure}

For illustration, we consider a random realization of the GBS horizontal locations as shown in Fig. \ref{trajectory}. Under this setup, we first obtain a maximum SNR target that can be achieved throughout the UAV mission by increasing $\bar{\rho}$ and checking the feasibility of Problem (P1), which is given by $\bar{\rho}_{\max}=14.69$dB. By considering this SNR target, i.e., $\bar{\rho}=\bar{\rho}_{\max}$, we show in Fig. \ref{trajectory} the proposed and the optimal trajectory designs, which are observed to be quite similar. Specifically, the GBS-UAV association sequences obtained by our proposed and the optimal designs are ${\mv{I}}=[1,10,11,6,8]^T$ and ${\mv{I}}=[1,10,9,2,11,6,8]^T$, respectively, whose corresponding handover locations are plotted in Fig. \ref{trajectory}. In addition, we illustrate the straight flight trajectory in Fig. \ref{trajectory}. It is observed that the given SNR target is not always attainable during the UAV mission with this trajectory, since a significant portion of this trajectory lies in the region where even the closest GBS has horizontal distance larger than $\bar{d}=\sqrt{\tfrac{\gamma_0}{\bar{\rho}}-(H-H_G)^2}$.

\begin{figure}[!htb]
  \centering
  \includegraphics[width=8cm]{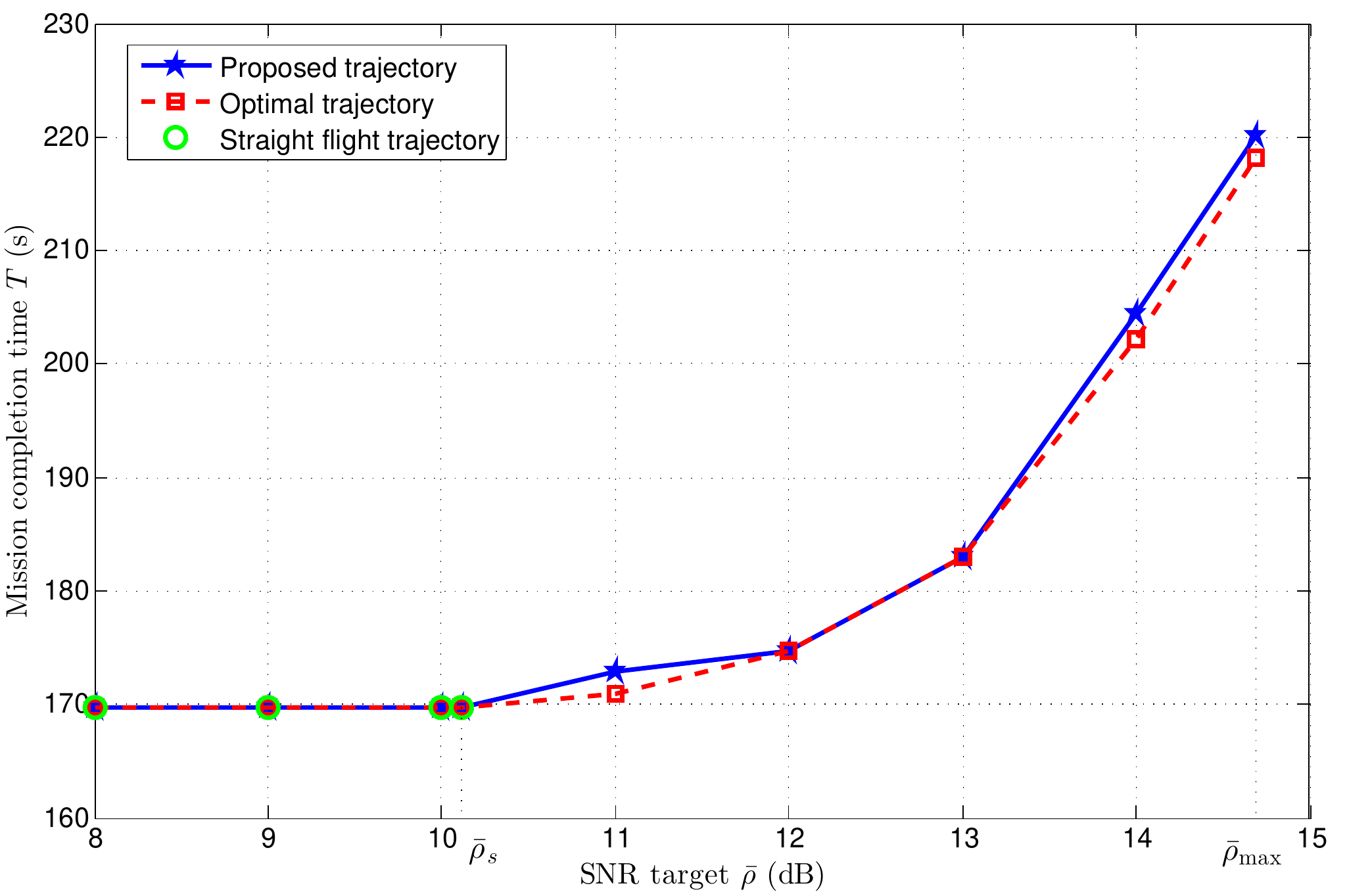}
  \vspace{-4mm}
  \caption{Mission completion time ${T}$ versus SNR target $\bar{\rho}$.}\label{time}
\vspace{-7mm}
\end{figure}

Moreover, we show in Fig. \ref{time} the mission completion time $T$ versus the SNR target $\bar{\rho}$ with our proposed trajectory design, the optimal trajectory design, and the straight flight trajectory, respectively. It is observed that although the straight flight trajectory achieves minimum mission completion time, it becomes infeasible as the SNR target exceeds a threshold given by $\bar{\rho}_s=10.12$dB, which is $4.57$dB lower than $\bar{\rho}_{\max}$, the maximum SNR target achievable by the other two trajectory designs. This thus validates the importance of trajectory design under the new connectivity constraint as investigated in this paper. On the other hand, it is observed that our proposed trajectory performs closely to the optimal trajectory for all SNR targets $\bar{\rho}$, yet with substantially reduced complexity required as discussed in Section \ref{sec_solution}. Furthermore, we randomly generate $500$ independent GBS locations, and evaluate the mission completion time required for our proposed and the optimal trajectory designs given the maximum achievable SNR target $\bar{\rho}=\bar{\rho}_{\max}$ under each setup. It is found that on average, our proposed design requires only $0.38\%$ more mission completion time compared to the optimal design, thus further validating the near-optimality of our proposed design.
\vspace{-6mm}
\section{Conclusion}
\vspace{-1mm}
This paper proposes an efficient UAV trajectory design solution for a cellular-enabled UAV communication system, where the UAV has a mission of flying between a pair of initial and final locations. Specifically, we formulate the UAV trajectory optimization problem to minimize the mission completion time, subject to a minimum received SNR constraint of the UAV-cellular communication link and a maximum UAV speed constraint. By applying graph theory and convex optimization techniques, we devise efficient algorithms for checking the feasibility and finding an approximate solution of the formulated problem. The proposed trajectory design is numerically verified to achieve near-optimal performance.

\appendices
\section{Proof of Proposition \ref{prop_feas}}\label{proof_prop_feas}
To start with, we prove the ``if'' part by showing that a feasible solution to Problem (P1-F) can be found with any given GBS-UAV association sequence $\mv{I}$ that satisfies the conditions in (\ref{feas_i})-(\ref{feas_I}). Specifically, we let the UAV fly from $U_0$ to $U_F$ following a path consisting of $N$ connected line segments characterized by $N+1$ discrete points with common altitude $H$, where the horizontal locations of the starting and ending points of each $i$th line segment are denoted as ${\mv{u}}^{i-1}$ and ${\mv{u}}^i$, respectively. The set $\{{\mv{u}}^i\}_{i=0}^N$ is given by ${\mv{u}}^0={\mv{u}}_0$, ${\mv{u}}^N={\mv{u}}_F$, and
\begin{align}
{\mv{u}}^i={\mv{g}}_{I_i}+\bar{d}\frac{{\mv{g}}_{I_{i+1}}-{\mv{g}}_{I_i}}{\|{\mv{g}}_{I_{i+1}}-{\mv{g}}_{I_i}\|},\quad i=1,...,N-1.
\end{align}
It can be shown from (\ref{feas_i}), (\ref{feas_f}) and (\ref{feas_GBS}) that
\begin{align}
&\|{\mv{u}}^i-{\mv{g}}_{I_i}\|\leq \bar{d},\quad i=1,...,N\label{if1}\\
&\|{\mv{u}}^i-{\mv{g}}_{I_{i+1}}\|\leq \bar{d},\quad i=0,...,N-1.\label{if2}
\end{align}
Therefore, for any point in the $i$th line segment with horizontal location ${\mv{u}}^i(p)=p{\mv{u}}^{i-1}+(1-p){\mv{u}}^{i},\ \forall p\in [0,1]$, we have
\begin{align}\label{u3}
&\|{\mv{u}}^i(p)-{\mv{g}}_{I_i}\|=\|p({\mv{u}}^{i-1}-{\mv{g}}_{I_i})+(1-p)({\mv{u}}^{i}-{\mv{g}}_{I_i})\|\nonumber\\
\!\!\overset{(A_1)}{\leq}\!&p\|{\mv{u}}^{i-1}\!-\!{\mv{g}}_{I_i}\|\!+\!(1-p)\|{\mv{u}}^{i}\!-\!{\mv{g}}_{I_i}\|\!\!\overset{(A_2)}{\leq}\!\! \bar{d}, i=1,...,N,
\end{align}
where $(A_1)$ is due to the triangle inequality, and $(A_2)$ is resulted from (\ref{if1}) and (\ref{if2}). It then follows from (\ref{u3}) that with the above UAV path and arbitrary UAV velocity $\dot{\mv{u}}(t)$, the resulting UAV trajectory satisfies the constraints in (\ref{P1c_i}), (\ref{P1c_f}) and (\ref{P1c_SNR}), which thus completes the proof of the ``if'' part.

On the other hand, we prove the ``only if'' part by showing that given any feasible solution $(T,{\mv{u}}(t))$ to Problem (P1-F), we can always construct $\{I_i\}_{i=1}^N$ that satisfies the conditions in (\ref{feas_i})-(\ref{feas_I}). Specifically, given any feasible solution $(T,{\mv{u}}(t))$ to Problem (P1-F), we can always find a finite number $N$ to divide $[0,T]$ into $N$ intervals and construct $\{I_i\}_{i=1}^N$, where $\arg\underset{m\in \mathcal{M}}{\min}\ \|{\mv{u}}(t)-{\mv{g}}_m\|=I_i$ and $\|{\mv{u}}(t)-{\mv{g}}_{I_i}\|\leq \bar{d}$ hold when $t$ lies in the $i$th interval, $\forall i\in\{1,...,N\}$. Note that the condition in (\ref{feas_I}) is automatically satisfied by $\{I_i\}_{i=1}^N$. Next, we construct $\{{\mv{u}}^i\}_{i=0}^N$ by defining ${\mv{u}}^0={\mv{u}}_0$ and letting ${\mv{u}}^i$ denote the horizontal location of the UAV at the end of the $i$th interval, $\forall i\in \{1,...,N\}$, with ${\mv{u}}^N={\mv{u}}_F$. It then follows that the constructed $\{I_i\}_{i=1}^N$ and $\{{\mv{u}}^i\}_{i=0}^N$ satisfy (\ref{if1}) and (\ref{if2}). As a result, it can be readily shown that $\{I_i\}_{i=1}^N$ satisfies the conditions in (\ref{feas_i}) and (\ref{feas_f}). Moreover, we have
\begin{align}
&\|{\mv{g}}_{I_{i+1}}-{\mv{g}}_{I_i}\|=\|({\mv{u}}^i-{\mv{g}}_{I_i})-({\mv{u}}^i-{\mv{g}}_{I_{i+1}})\|\nonumber\\
\!\!\overset{(A_3)}{\leq}\!&\|{\mv{u}}^i-{\mv{g}}_{I_i}\|\!+\!\|{\mv{u}}^i-{\mv{g}}_{I_{i+1}}\|\!\overset{(A_4)}{\leq}\! 2\bar{d},\  i=1,...,N-1,\label{onlyif3}
\end{align}
where $(A_3)$ is due to the triangle inequality, and $(A_4)$ results from (\ref{if1}) and (\ref{if2}). Hence, the constructed $\{I_i\}_{i=1}^N$ also satisfies the condition in (\ref{feas_GBS}), which thus completes the proof of the ``only if'' part.
\section{Proof of Proposition \ref{prop_P2eq}}\label{proof_prop_P2eq}
First, given any feasible solution $(T,{\mv{u}}(t),{\mv{I}},\{T_i\}_{i=1}^N)$ to Problem (P2), it follows from (\ref{P2c_SNR}) that $\underset{m\in\mathcal{M}}{\min}\ \|{\mv{u}}(t)-{\mv{g}}_m\|\leq \|{\mv{u}}(t)-{\mv{g}}_{I_i}\|\leq \bar{d}$ holds for any $t\in\left[\sum_{j=1}^{i-1}T_j,\sum_{j=1}^i T_j\right]$ and $i=1,...,N$. Thus, $(T,{\mv{u}}(t))$ is a feasible solution to Problem (P1) and achieves the same objective value as Problem (P2) with the solution $(T,{\mv{u}}(t),{\mv{I}},\{T_i\}_{i=1}^N)$. Hence, the optimal value of Problem (P1) is no larger than that of Problem (P2). On the other hand, for any given feasible solution $(T,\mv{u}(t))$ to Problem (P1), we can always divide $[0,T]$ into $N$ intervals denoted by $\left[\sum_{j=1}^{i-1}T_j, \sum_{j=1}^i T_j\right],\ i=1,...,N$, such that $I_i=\arg\underset{m\in\mathcal{M}}{\min}\ \|{\mv{u}}(t)-{\mv{g}}_m\|$ and $\|{\mv{u}}(t)-{\mv{g}}_{I_i}\|\leq \bar{d}$ hold for any $t\in\left[\sum_{j=1}^{i-1}T_j,\sum_{j=1}^i T_j\right]$ and $i=1,...,N$. By following similar procedure as in the ``only if'' part of the proof of Proposition \ref{prop_feas}, it can be shown that ${\mv{I}}=[I_1,...,I_N]^T$ satisfies the constraints in (\ref{feas_i})-(\ref{feas_I}). Hence, $(T,{\mv{u}}(t),{\mv{I}},\{T_i\}_{i=1}^N)$ can be shown to be feasible for Problem (P2) and achieves the same objective value as Problem (P1) with the solution $(T,{\mv{u}}(t))$. The optimal value of Problem (P2) is thus no larger than that of Problem (P1). Therefore, Problem (P1) and Problem (P2) have the same optimal value, which completes the proof of Proposition \ref{prop_P2eq}.
\section{Proof of Proposition \ref{prop_P2}}\label{proof_prop_P2}
We prove Proposition \ref{prop_P2} by showing that for any feasible solution to Problem (P2) denoted by $(\tilde{T},\tilde{\mv{u}}(t),{\mv{I}},\{\tilde{T}_i\}_{i=1}^N)$, we can always construct a feasible solution to Problem (P2) denoted by $({T},{\mv{u}}(t),{\mv{I}},\{T_i\}_{i=1}^N)$ that satisfies the conditions in (\ref{Ti}), (\ref{ut}) and (\ref{T}), and achieves no larger objective value of Problem (P2) compared to $(\tilde{T},\tilde{\mv{u}}(t),{\mv{I}},\{\tilde{T}_i\}_{i=1}^N)$. We start by constructing the same handover locations in ${\mv{u}}(t)$ as those in $\tilde{\mv{u}}(t)$, i.e., ${\mv{u}}^i=\tilde{\mv{u}}\left(\sum_{j=1}^i \tilde{T}_j\right),\ i=0,...,N$. Then, note that $\tilde{T}_i$ denotes the time duration for the UAV to fly from ${\mv{u}}^{i-1}$ to ${\mv{u}}^i$, thus $\tilde{T}_i\geq \frac{\|{\mv{u}}^i-{\mv{u}}^{i-1}\|}{V_{\max}},\ i=1,...,N$ should hold, since $\|{\mv{u}}^i-{\mv{u}}^{i-1}\|$ is the minimum distance between ${\mv{u}}^{i-1}$ and ${\mv{u}}^i$, and $V_{\max}$ is the maximum allowable speed. By noting that $T_i=\frac{\|{\mv{u}}^i-{\mv{u}}^{i-1}\|}{V_{\max}}$ holds as shown in (\ref{Ti}), we have $T_i\leq \tilde{T}_i,\ i=1,...,N$, and consequently $\tilde{T}=\sum_{i=1}^N\tilde{T}_i\geq T=\sum_{i=1}^N{T}_i$. The proof of Proposition \ref{prop_P2} is thus completed.
\section{Proof of Lemma \ref{lemma_I}}\label{proof_lemma_I}
Consider a feasible solution of ${\mv{I}}$ to Problem (P3) given by $\hat{\mv{I}}=\left[\hat{I}_1,...,\hat{I}_k,...,\hat{I}_q,...,I_{\hat{N}}\right]^T$, where $\hat{I}_k=\hat{I}_q$, and another feasible solution of $\mv{I}$ by removing the $(k+1)$th to the $q$th elements in $\hat{\mv{I}}$, which is given by $\tilde{\mv{I}}=\left[\hat{I}_1,...,\hat{I}_k,\hat{I}_{q+1},...,\hat{I}_{\hat{N}}\right]^T$. It can be shown that for given $\hat{\mv{I}}$ and any feasible $\{\hat{\mv{u}}^i\}_{i=0}^{\hat{N}}$ to Problem (P3), the resulting objective value is given by $\hat{s}\overset{\Delta}{=}\sum_{i=1}^{k-1} \|\hat{\mv{u}}^i-\hat{\mv{u}}^{i-1}\|+\sum_{i=k}^q \|\hat{\mv{u}}^i-\hat{\mv{u}}^{i-1}\|+ \sum_{i=q+1}^{\hat{N}}\|\hat{\mv{u}}^i-\hat{\mv{u}}^{i-1}\|$. On the other hand, it can be shown that $(\tilde{\mv{I}},\{\tilde{\mv{u}}^i\}_{i=0}^{\hat{N}-(q-k)})$ with $\tilde{\mv{u}}^i=\hat{\mv{u}}^i,\ i=0,...,k-1$ and $\tilde{\mv{u}}^i=\hat{\mv{u}}^{i+(q-k)},\ i=k,...,\hat{N}-(q-k)$ is also a feasible solution to Problem (P3), whose resulting objective value is given by $\tilde{s}\overset{\Delta}{=}\sum_{i=1}^{k-1} \|\hat{\mv{u}}^i-\hat{\mv{u}}^{i-1}\|+\|\hat{\mv{u}}^q-\hat{\mv{u}}^{k-1}\|+\sum_{i=q+1}^{\hat{N}}\|\hat{\mv{u}}^i-\hat{\mv{u}}^{i-1}\|$.
By applying the triangle inequality, it can be shown that $\|\hat{\mv{u}}^q-\hat{\mv{u}}^{k-1}\|=\|\sum_{i=k}^q(\hat{\mv{u}}^i-\hat{\mv{u}}^{i-1})\|\leq \sum_{i=k}^q\|\hat{\mv{u}}^i-\hat{\mv{u}}^{i-1}\|$ holds. It then follows that $\tilde{s}\leq \hat{s}$ holds, i.e., the objective value of Problem (P3) with the solution $(\tilde{\mv{I}},\{\tilde{\mv{u}}^i\}_{i=0}^{\hat{N}-(q-k)})$ is no larger than that of Problem (P3) with the solution $(\hat{\mv{I}},\{\hat{\mv{u}}^i\}_{i=0}^{\hat{N}})$. Therefore, the optimal solution to Problem (P3) should satisfy $I_i\neq I_j,\ \forall i\neq j$, and thus the length of the optimal $\mv{I}$ should not exceed the total number of GBSs, i.e., $N\leq M$ should hold. This completes the proof of Lemma \ref{lemma_I}.
\bibliographystyle{IEEEtran}
\bibliography{CellularUAV}
\end{document}